\documentclass[usenatbib]{mn2e}
\usepackage{natbibmnfix, graphicx, times, amsmath, epsfig, amssymb}
\usepackage{amsmath}
\usepackage{verbatim}
\newcommand{\zpeak}{z_{\rm peak}}

\newcommand{\cmfast}{\textsc{\small 21CMFAST}}

\newcommand{\Ts}{T_{\rm S}}

\newcommand{\nf}{x_{\rm HI}}
\newcommand{\avenf}{\bar{x}_{\rm HI}}

\newcommand{\lya}{Ly$\alpha$}

\newcommand{\Msun}{M_\odot}
\newcommand{\Tvir}{T_{\rm vir}}

\newcommand{\Tcmb}{T_\gamma}
\newcommand{\delT}{\delta T_b}

\newcommand{\delNL}{\delta_{\rm nl}}

\newcommand\lsim{\mathrel{\rlap{\lower4pt\hbox{\hskip1pt$\sim$}}
        \raise1pt\hbox{$<$}}}
\newcommand\gsim{\mathrel{\rlap{\lower4pt\hbox{\hskip1pt$\sim$}}
        \raise1pt\hbox{$>$}}}
\def\myputfigure#1#2#3#4#5%
{\vskip#5pt\makebox[0pt]{\hskip#2in
\includegraphics[width=#3\textwidth]{#1}}\vskip#4pt\hfill}

\begin{document}

\title[The X-ray spectra of the first galaxies: 21cm signatures]
      {The X-ray spectra of the first galaxies: 21cm signatures}
\author[F. Pacucci et al.]
{Fabio Pacucci$^1$ \thanks{fabio.pacucci@sns.it},
Andrei Mesinger$^1$, Stefano Mineo$^2$, Andrea Ferrara$^{1,3}$ \\
$^1$Scuola Normale Superiore, Piazza dei Cavalieri, 7  56126 Pisa, Italy \\
$^2$Harvard-Smithsonian Center for Astrophysics, 60 Garden Street Cambridge, MA 02138, USA \\
$^3$Kavli Institute for the Physics and Mathematics of the Universe (WPI), Todai Institutes for Advanced Study, the University of Tokyo \\
}
             
\date{submitted to MNRAS}

\maketitle
             
\begin{abstract}
The cosmological 21cm signal is a physics-rich probe of the early Universe, encoding information about both the ionization and the thermal history of the intergalactic medium (IGM).  The latter is likely governed by X-rays from star-formation processes inside very high redshift ($z \gsim 15$) galaxies. 
Due to the strong dependence of the mean free path on the photon energy, the X-ray SED can have a significant impact on the interferometric signal from the cosmic dawn.
Recent \textit{Chandra} observations of nearby, star-forming galaxies show that their SEDs are more complicated than is usually assumed in 21cm studies.
In particular, these galaxies have ubiquitous, sub-keV thermal emission from the hot interstellar medium (ISM), which generally dominates the soft X-ray luminosity (with energies $\lsim$ 1 keV, sufficiently low to significantly interact with the IGM).
Using illustrative soft and hard SEDs, we show that 
the IGM temperature fluctuations in the early Universe would be substantially increased if the X-ray spectra of the first galaxies were dominated by the hot ISM, compared with X-ray binaries with harder spectra.
The associated large-scale power of the 21cm signal would be higher by a factor of $\sim$ three.
More generally, we show that the peak in the redshift evolution of the large-scale ($k \sim 0.2$ $\mathrm{Mpc}^{-1}$) 21cm power is a robust probe of the soft-band SED of the first galaxies, and importantly, is not degenerate with their bolometric luminosities.  On the other hand, the redshift of the peak constrains the X-ray luminosity and halo masses which host the first galaxies. 

\end{abstract}


\setcounter{footnote}{1}

%%%%%%%%%%%%%%%%%%%%%%%%%%%%%%%%%%%%%%%%%%%%%%%%%%%%%%%%%%%%%%%%%%%%%%
%% SECTION 1: INTRODUCTION
%%%%%%%%%%%%%%%%%%%%%%%%%%%%%%%%%%%%%%%%%%%%%%%%%%%%%%%%%%%%%%%%%%%%%%
\section{Introduction}
\label{sec:intro}
%The formation of the first stars and compact objects of our Universe is one of the primary topics in the modern cosmological research. The redshifted 21cm line from neutral hydrogen will arguably provide the most important insights into these epochs. 
The redshifted 21cm line is sensitive to the thermal and ionization state of the cosmic gas, making it a powerful probe of the early Universe. As it is a line transition, it has the potential to map out the three dimensional structure of cosmic gas and its evolution.   First generation interferometers, including the Low Frequency Array (LOFAR; \citealt{vanHaarlem13})\footnote{http://www.lofar.org/},  Murchison Wide Field Array (MWA; \citealt{Tingay12})\footnote{http://www.mwatelescope.org/}, and the Precision Array for Probing the Epoch of Reionization (PAPER; \citealt{Parsons10})\footnote{http://eor.berkeley.edu}, are already taking data. Their focus is on a statistical detection of reionization, though even earlier epochs of heating (when the cosmic gas was heated to temperatures above the CMB) could be detectable \citep{ME-WH14}.  Second generation instruments, like the Square Kilometre Array (SKA; \citealt{SKA12})\footnote{http://www.skatelescope.org/} will be coming on-line soon, with high sensitivity and wide frequency coverage, allowing us to witness the birth of the very first galaxies through their imprint on the intergalactic medium (IGM).

X-rays play a very important role during these epochs. Reionization with a significant X-ray contribution proceeds more uniformly,
% resulting in a suppression of large-scale 21cm fluctuations later than predicted by standard UV-driven reionization models
complicating the interpretation of 21cm fluctuations on large-scales \citep{MFS13}.   More importantly, X-rays are thought to be responsible for heating the IGM to temperatures above the CMB, before reionization gets well-underway (e.g. \citealt{Furlanetto06, MO12}).  In fiducial models, the large-scale temperature fluctuations during this heating epoch are responsible for the strongest 21cm interferometric signal, an order of magnitude greater than the signal during reionization. Understanding the timing and homogeneity of X-ray heating is critical in interpreting 21cm observations of the pre-reionization and reionization epochs (e.g. \citealt{PF07, ME-WH14}).

%In the early Universe X-rays play an important role, likely dominating its thermal evolution and hence the pre-reionization epoch, with $10<z<20$.
%The early Universe is dominated by X-rays and it is highly likely that the peak in the amplitude of large-scale fluctuations of the 21cm signal occurred during
%the pre-reionization epoch, when X-rays began heating the cold IGM. Sourced by strong absorption of cold gas against the CMB and large temperature fluctuations in the IGM, the 21cm power during X-ray heating is expected to be at least an order of magnitude higher than that during reionization (e.g. \citealt{PF07, MF07, Baek10, MFS13}).

A common approach is to parameterize our uncertainty of the early X-ray background by fixing the galactic X-ray spectral energy distribution (SED), and varying its normalization, i.e. bolometric luminosity (e.g. \citealt{Furlanetto06, Santos11, CL13, ME-WH14}; though see also exploratory work in \citealt{PF07, Baek10, MFS13}).  The X-ray luminosity of the first galaxies\footnote{For convenience, we use the adjective ``first'' somewhat imprecisely, referring to the galaxies responsible for heating the IGM, which likely occurs at $z\sim$10--20 (see below).  The very first galaxies could appear even earlier ($z\sim30$), though star formation inside these rare mini-halos is likely insufficient to significantly heat the IGM (e.g. \citealt{MO12}).  Nevertheless, our qualitative conclusions are not affected by the precise redshift at which the relevant galaxies appear (see below).} regulates the timing of the heating epoch. 
 However, the actual X-ray SED should also be important in setting the signal, as the mean free path of X-rays through the IGM, $\lambda_{\rm X}$, has a very strong dependence on the photon energy (e.g. \citealt{FOB06, McQuinn12}):
\begin{equation}
\label{eq:mfp}
\lambda_{\rm X} \approx 34 ~ \avenf^{-1} \left( \frac{E_{\rm X}}{\rm 0.5 ~ keV} \right)^{2.6} \left( \frac{1+z}{15} \right)^{-2} ~ {\rm comoving ~ Mpc} ~,
\end{equation}
where $\avenf$ is the mean neutral fraction of the IGM.
Soft photons are much more likely to be absorbed closer to the galaxies, while high energy photons heat (or ionize) the IGM more uniformly. Indeed, \citet{MFS13} showed that if X-ray heating is dominated by high-energy photons, the redshift evolution of the amplitude of the large-scale 21cm power spectrum does not show an associated pronounced peak.  It is important to also note that because of this strong energy dependence of $\lambda_{\rm X}$, photons with energies $\gsim$ 2 keV effectively free-steam, barely interacting with the IGM; this makes the soft X-ray SED much more relevant for the 21cm signal.

Observations show that the SED of local galaxies is more complicated than is usually assumed in 21cm studies.  Locally, the hot ISM contributes significantly to the galaxy's soft X-ray emission (e.g. \citealt{Strickland_2000, Grimes05, Owen09, Strickland_2004, Li_Wang_2013}; see the review in \S 7.1 of \citealt{Mineo_ISM}).  As an example, we note that using \textit{Chandra}, \citet{Mineo_ISM} recently studied the diffuse emission in a local sample of 21 star-forming galaxies, finding sub-keV thermal emission from the hot ISM in {\it every} galaxy in the sample.  The stacked, bolometric soft-band (0.5--2 keV) luminosity per star formation rate (SFR) of the thermal emission is comparable to that from resolved sources, dominated by high mass X-ray binaries (HMXBs) with much harder spectra (e.g. \citealt{GGS04, Mineo_HMXB}).
%A softer SED could was which can dominate the total luminosity in the relevant soft bands (with energies $<(1-2) \, \mathrm{keV}$, sufficiently low to actually interact with the IGM). These would have short mean free paths, and the corresponding SEDs would result in strong temperature fluctuations in the early universe, with regions near galaxies much hotter than the distant IGM.  Strong temperature fluctuations could boost the 21cm interferometric signal, making it detectable with even first generation instruments (see \citealt{ME-WH14}).

In this paper we illustrate the impact of the X-ray SED of the first galaxies on the 21cm power spectrum.  We use simple models representative of  dominant populations of either soft (corresponding to the hot ISM) or hard (corresponding to HMXBs) X-ray sources.  To show the robustness of our results, we also vary the X-ray luminosity per SFR (SED normalization) and the halo mass which hosts the dominant galaxy population.
%explore the effect of additional astrophysical parameters, such as the virialization temperature $T_{vir}$ (which controls the time when the main sources of high-energy photons appear) and the galactic X-ray emissivity $\zeta_X$, in order to decouple these effects from the one driven by the SED.

As this work was nearing completion, a related study was published by \citet{FBV14}. The most important distinction between the two works is that our analysis is motivated by {\it Chandra} observations of nearby star-forming galaxies, rather then a theoretical model of HMXBs.  Furthermore, our proof-of-concept focuses on predicting qualitative trends which are robust to the many astrophysical uncertainties.

This paper is organized as follows. In \S \ref{sec:SED} we discuss possible contributions to the X-ray SED of high$-z$ galaxies, placing them in the context of recent \textit{Chandra} observations. 
 In \S \ref{sec:21cm} we present our simulations of the
 cosmological 21cm signal. In \S \ref{sec:results} we discuss our main results, showing how the SED has a robust imprint in the 21cm signal.  Finally, we conclude in \S \ref{sec:conc}.
Unless stated otherwise, we quote all quantities in comoving units.
Throughout, we adopt recent Planck cosmological parameters \citep{Planck_Parameters}: $(\Omega_m, \Omega_{\Lambda}, \Omega_b, h, n_s, \sigma_8)= (0.32, 0.68, 0.049, 0.67, 0.96, 0.83)$.

%%%%%%%%%%%%%%%%%%%%%%%%%%%%%%%%%%%%%%%%%%%%%%%%%%%%%%%%%%%%%%%%%%%%%%
%% SECTION 2
%%%%%%%%%%%%%%%%%%%%%%%%%%%%%%%%%%%%%%%%%%%%%%%%%%%%%%%%%%%%%%%%%%%%%%
\section{X-rays from the first galaxies}
\label{sec:SED}

As we do not know the X-ray SEDs of high-redshift, $z>10$, galaxies\footnote{QSOs at $z\lsim6$ have been detected in X-rays (e.g. \citealt{Brandt_1999, Fan_1999}), as well as galaxies at $z\lsim4$ through stacking analysis \citep{Basu_Zych_2013}.}
we are forced to make educated guesses, motivated by observations of low-z ($z\lsim4$) galaxies.  In the local Universe, active galactic nuclei (AGN) dominate the X-ray background (XRB; e.g. \citealt{Moretti12}).  However at high-redshifts ($z\gsim5$; e.g. \citealt{HM12, Fragos13}), the contribution of AGN to the X-ray background should become sub-dominant to that of end products of stellar evolution, accreting gas from companion stars.  These are characterized by the masses of their donor stars, and comprise HMXBs, intermediate mass X-ray binaries (IMXBs), low mass X-ray binaries (LMXBs), cataclysmic variables and active binaries.  However, the characteristic timescales of all but the HMXBs are longer than the Hubble time at the very high redshifts of interest.  Furthermore, the bolometric\footnote{We stress again that high-energy X-rays are unlikely to interact with the IGM at redshift relevant for the 21cm signal, given their long mean free paths. Hence the soft-band ($\lsim2$ keV) SED and luminosity is more relevant for predicting the 21cm signal.}
X-ray luminosity of local, star-forming galaxies is found to be dominated by resolved HMXBs (e.g. \citealt{Grimm_2003, Ranalli_2003, Swartz_2004, Persic_2007, Lehmer_2010, Swartz_2011, Walton_2011, Mineo_HMXB}).
 For these reasons, many studies of early IGM heating focus on HMXBs as the primary source of X-rays in star-forming galaxies (e.g. \citealt{Furlanetto06, PF07, MFC11, Santos11}; though see, e.g., \citealt{Valdes13} and Evoli et al., in prep. for more exotic models in which heating can be dominated by annihilating dark matter).

On the other hand, the hot ISM could contribute a significant amount of soft X-rays.  Heated by supernovae (SNe) explosions and winds to temperatures of 10$^{6-7}$ K, this hot plasma emits X-rays through a combination of thermal bremsstrahlung and metal line cooling.  It is diffuse and more spatially extended than the point sources discussed above.  Its presence is typically detected in normal star-forming galaxies and starbursts (e.g \citealt{Strickland_2000, Strickland_2004_2, Grimes_2005, Mineo_ISM, Li_Wang_2013}), as well as in high-resolution ISM simulations of the first, atomically-cooled galaxies (e.g. \citealt{Wise_Abel_2012}; Aykutalp et al., in prep).
  The contribution of soft emission from the hot ISM to the X-ray heating epoch has not been considered previously.

Below we take HMXBs and the hot ISM as the two potential sources of X-ray emission from the first galaxies.  In nearby galaxies, the total luminosity of both of these sources is observed to be proportional to the galaxy's star-formation rate (e.g. \citealt{GGS04}).  Interestingly, {\it both HMXBs and the hot ISM have a comparable, observed soft-band (0.5--2 keV) luminosity per SFR}: $\sim$ 8 and 5 $\times10^{38}$ erg s$^{-1}$ $\Msun$ yr$^{-1}$, respectively \citep{Mineo_HMXB, Mineo_ISM}.  However, {\it their SEDs are dramatically different}, as we discuss further below.  We also make the distinction between the {\it intrinsic} and {\it observed (or emerging)} SEDs, the later including absorption from the host galaxy.

\begin{figure}
\vspace{-1\baselineskip}
{
\includegraphics[width=0.5\textwidth]{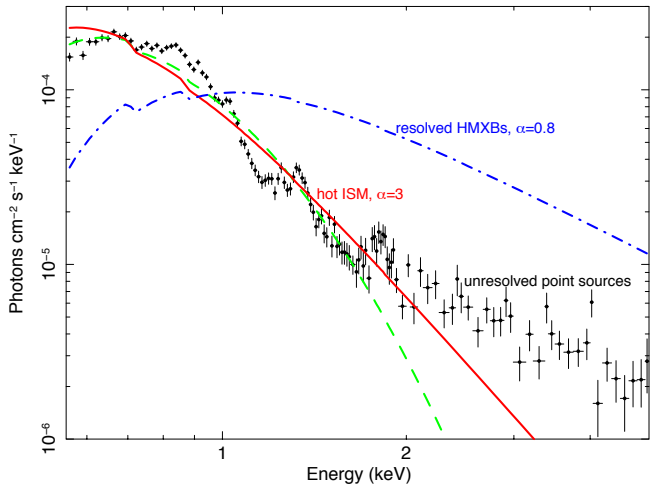}
}
\caption{
Composite observed X-ray SEDs from a sample of 21 local, star-forming galaxies (for further details, see \citealt{Mineo_ISM}). Points correspond to unresolved emission, which at low energies ($\lsim 2$ keV) is dominated by the hot ISM, while at high energies ($\gsim 2$ keV) by faint point sources (unresolved HMXBs, LMXBs, cataclysmic variables and active binaries).  A best-fit thermal bremsstrahlung profile for the hot ISM is shown with the green, dashed curve.  
Our fiducial toy-model SED for the hot ISM (a power-law with energy index $\alpha$=3) is indicated with a solid red curve, which is a good match to the bremsstrahlung profile. The intrinsic SED for resolved HMXBs (blue dot-dashed curve) is instead a power-law with energy index $\alpha$=0.8, based on the average spectrum of HMXBs (see \citealt{Swartz04, Mineo_HMXB}).
%The intrinsic emission associated with the red solid (blue dot-dashed) curves represent our soft (hard) model SEDs, which  are a good match to the profiles of the hot ISM (HMXBs); see text for details. 
 The normalization of the blue dot-dashed curve is done to preserve the observed relative ratio of the soft-band (0.5--2 keV) luminosities per unit SFR for the HMXB and hot ISM \citep{Mineo_HMXB, Mineo_ISM}.
% We caution that the points include a deconvolution of the {\it Chandra} instrument response, which is not standard practice when presenting X-ray observations.
The red (blue) curve corresponds to our fiducial soft (hard) model for the intrinsic emission.
  In order to compare to the observed emergent spectra, all curves include intrinsic absorption by an equivalent HI column density of $N_{\rm HI}\sim$2--3$\times10^{21}$ cm$^{-2}$, assuming solar abundances and metallicity.}
\label{fig:spectrum}
\vspace{-0.5\baselineskip}
\end{figure}

\subsection{The intrinsic SED of the hot, diffuse ISM}

Star-forming galaxies are known to output abundant X-ray emission from hot ionized gas of sub-keV temperatures, whose luminosity correlates with the SFR of the host galaxy (e.g. \citealt{Strickland_2000, Strickland_2004_2, Grimes_2005, Li_Wang_2013}). Recently, \citet{Mineo_ISM} presented Chandra observations of 21 local, star-forming galaxies, isolating the contribution of hot, diffuse ISM. They took special care of various systematic effects and controlled the contamination of unresolved emission by bright compact sources of all types as well as by unresolved faint high-mass X-ray binaries. Every galaxy in their sample showed evidence of hot plasma, with temperatures in the range of 0.2--0.3 keV, with 1/3 of their sample showing evidence of a second thermal peak at $\sim$0.7 keV. 

In Fig. \ref{fig:spectrum}, we show the composite SED of the unresolved emission from their sample of 21 galaxies, corrected for the \textit{Chandra} instrument response (black points).  At these temperatures the emission from the hot plasma should be governed by metal line cooling and thermal bremsstrahlung (e.g. \citealt{GS83}).  At soft energies ($\lsim$ 2 keV), the composite can be roughly approximated by either templates: (i) a solar-metallicity, thermal (\textsc{\small MEKAL} from \textsc{\small XSPEC}) component with mean energy $\langle$kT$\rangle\sim$ 0.3 keV; or (ii) thermal bremsstrahlung, also with $\langle$kT$\rangle\sim$ 0.3 keV.  The latter is shown with the green dashed curve in Fig. \ref{fig:spectrum}.
 At higher energies ($\gsim$ 2 keV), the composite SED of the unresolved emission is dominated by faint point sources with harder spectra.

With the red solid curve we also show in Fig. \ref{fig:spectrum} a power-law SED, with the specific luminosity scaling as $L_{\rm X} \propto E_{\rm X}^\alpha$ and an energy index of $\alpha=3$.  In our analysis below, we adopt this simple power-law for the intrinsic emission for our fiducial soft SED, as it provides a reasonable fit to the SED of the hot, diffuse ISM.

%Local star-forming galaxies are also known to possess significant amounts of hot ionized gas of sub-keV temperatures, which is a source of copious X-ray emission (\citealt{Grimes05, Owen09}). The morphology of the diffuse X-ray emission suggests \citep{Strickland2000} that the gas is in the state of outflow, driven by the effect of SNe and winds from massive stars.

%Composite spectra from \cite{Mineo_ISM} show a thermal peak at $<kT_1>=0.24 \, \mathrm{keV}$ for the entire sample of galaxies and a second peak at $<kT_2>=0.71 \,\mathrm{keV}$ for about $1/3$ of them. It is important to note that the luminosity in the soft band of the diffuse ISM emission is comparable to the luminosity of HMXBs and surpasses it at energies $\gsim 1.5 \, \mathrm{keV}$. The total luminosity in the soft band is $L^x_{0.5-2.0 \, \mathrm{keV}} \approx 0.08 \times 10^{40} \, \mathrm{erg/s}$. The overall soft X-ray emission from local star-forming galaxies includes the contribution of both the ISM and the thermal part of the collective emission of faint and unresolved HMXBs (see Fig. 4 in \citealt{Mineo_ISM}). The right panel of Fig. \ref{fig:spectrum} shows the combination of all the spectra in Fig. 5 in \cite{Mineo_ISM}, deconvolved from the detector response, along with its best-fitting model.

\subsection{The intrinsic SED of HMXBs}

Composite SEDs of HMXBs generally follow a hard power-law, with a spectral energy index of $\alpha\approx$0.7--1 \citep{RGP95, Swartz04, Mineo_HMXB}. This corresponds to the so-called, 'hard state' resulting from Comptonization of soft photons on hot electrons ("corona"), in the vicinity of the compact object.
% inverse Compton scattering in the corona of the accreting black hole. 
Accreting compact objects  also show evidence of a 'soft state', which is believed to originate in the optically thick/geometrically thin accretion disk \citep{SS73}, well represented by a superposition of black body spectra with
% Some individual objects also show evidence of a 'soft state' with a black body spectrum with
 $\langle$kT$\rangle\lsim$ 1 keV
%, thought to stem from the accretion disk itself
 (for a review see \citealt{McClintock06} or \citealt{Gilfanov10}).
%  However, the prevalence of the 'soft state' is debatable, due to the increased difficulty of observations in this energy range and the  possible contamination by unresolved supernovae remnants \citep{Mineo_ISM}.
However, the composite SEDs of {\it bright} X-ray compact sources associated with young stellar populations, such as resolved HMXBs \citep{Mineo_HMXB} or ultra-luminous X-ray sources (ULXs; \citealt{Swartz04}) are typically dominated by the 'hard state' and are well fit with an absorbed power-law (with spectral energy index of $\sim 0.7-1$).\footnote{There is evidence that bright UXLs have a spectral cut-off at high-energies, $E_{\rm X}\gsim$ 6--10 keV (e.g. \citealt{Miayawaki09, Bachetti13}).  Our results are not sensitive to such modifications of the fiducial SED, since photons with energies above $E_{\rm X}\gsim 1.8 \avenf^{1/3} [(1+z)/15]^{1/2}$ keV have mean free paths exceeding the Hubble length (e.g. \citealt{McQuinn12}), and thus do not interact significantly with the IGM.  However, fiducial normalizations of the X-ray luminosity to SFR ratios (see below), as well as estimates of their redshift evolution, should account for possible high-energy cut-offs (e.g. \citealt{Kaaret_2014}).}

%Nonetheless, it seems that all the bright ULXs that have been observed with XMM with enough S/N show a curved spectrum instead of the canonical hard state. All those that are being observed with the NuSTAR observatory (which is sensitive to harder X-ray band) confirm the presence of such a curvature in their spectra. There is a growing X-ray community trying to push the idea of \textit{all} ULXs being characterized by these curved spectra. However: (i) we are considering all HMXBs, not only ULXs, and there is enough literature supporting our model for hard SED; (ii) it seems that not all the ULXs have such a curved spectrum, but only those powered by a stellar mass black hole accreting at super-Eddington rates. For example, the source HLX-1 \citep{Farrell_2009} has a peak luminosity of $\sim 10^{42} \mathrm{erg \,s ^{-1}}$ and it does not show any hint for a curved spectrum, due to the fact that it is a strong candidate for an IMBH accreting at sub-Eddington rates.

For our fiducial hard SED below, corresponding to the intrinsic emission from bright HMXBs, we adopt a power-law with energy index $\alpha=0.8$.  We show this hard power-law as a blue dot-dashed curve in Fig. \ref{fig:spectrum}, obscured with an equivalent HI column density of $N_{\rm HI}\sim3\times10^{21}$ cm$^{-2}$ \citep{Mineo_HMXB}.

\subsection{Intrinsic absorption from the host galaxy}

The shape of the spectrum at low energies can depend strongly 
on the intrinsic absorption of the host galaxy.  For our purposes, it is useful to decompose the intrinsic absorption into two contributions: (i) an HI column density, and (ii) metal abundance.

For gas with a solar abundance and metallicity, the metals dominate the absorption of photons with energies $\gsim 0.5$ keV, while helium dominates at lower energies \citep{MM83}.  The standard approach in X-ray studies is to assume a solar abundance and metallicity when constructing the absorption profile, and then quote the associated HI column density ($N_{\rm HI}\sim$2--3$\times10^{21}$ cm$^{-2}$ in the case of the curves shown in Fig. \ref{fig:spectrum}).  However, the early galaxies at $z\sim$10--20 which drive the IGM heating should be much less enriched than local ones.  If the average sight-line out of the galaxy contains less metals, more soft photons would escape into the IGM for a given HI column density.  Just based on qualitative arguments about metal evolution, one would expect the emergent spectra of the first galaxies to be softer than shown in Fig. \ref{fig:spectrum}.

It is even less clear how to estimate the HI column density of the first galaxies.  Empirical trends suggest that the fraction of ionizing photons escaping galaxies increases rapidly towards high redshifts (e.g. \citealt{HM12, KF-G12}), perhaps driven by the shallower potential wells of typical galaxies which make the gas distribution more susceptible to feedback (e.g. \citealt{FL12, AFT12}).  If confirmed, it would be reasonable to assume that the average column density relevant for X-ray absorption follows similar qualitative trends.

 There could also be a relative difference in opacities for our two sources of X-rays: hot ISM and HMXBs.  One could imagine that the diffuse, spatially-extended hot ISM might have lower covering fractions of HI, compared to the HMXBs which can be embedded in dust clouds (which further attenuate soft X-rays).  Indeed, observations of the diffuse, hot ISM show that less than half of local, star-forming galaxies have
evidence of {\it any} host galaxy absorption \citep{Mineo_ISM}.

The discussion above is clearly very speculative, and depends on the unknown morphology and enrichment of the first galaxies.  The absorption of X-ray photons by the high-$z$ host galaxies could have a strong impact on the emerging X-ray SED, and we will return to this in future work.  For this proof-of-concept, we assume a very simple model for host-galaxy obscuration, truncating the intrinsic SED at energies below $<0.3$ keV.  Photons below this energy have an optical depth greater than unity, for the column densities greater than $N_{\rm HI}\gsim 10^{21.5}$ cm$^{-2}$, assuming that helium dominates the total opacity\footnote{We caution that metal abundance and column density are not disentangled in present analysis of local galaxies, which assumes solar abundances and metallicity.  If the sightline-averaged metallicity of the host galaxy is lower, the best fit HI column density would be higher, and vice versa.}.  In other words, we assume that the first galaxies have similar column densities as local ones, but take the gas to have a low metal abundance.  For illustration, we also present an 'extreme' model in which we truncate all photons with energies below $\leq1$ keV.  This model is 'extreme' in the sense that it assumes the first galaxies were much more obscured than even the more massive, evolved local ones (Fig. \ref{fig:spectrum}).

\subsection{Normalizations of the model SEDs}

Following previous work, we normalize our fiducial soft and hard SEDs with the parameter, $f_X\equiv N_X/0.1$, where $N_X$ is the total number of X-ray photons escaping the galaxy per stellar baryon.  For both our soft and hard spectra, the fiducial choice of $f_X = 1$ yields band-integrated luminosities per unit SFR comparable to observed ones.  Our hard SED yields a 0.5--8 keV luminosity of $L_{\rm X, 0.5-8keV}^{\alpha=0.8} \approx 5 \times 10^{39}$ erg s$^{-1}$ $\Msun$ yr$^{-1}$,
similar to the observed value of $L_{\rm X, 0.5-8keV}^{\rm obs, HMXB} \approx 3 \times 10^{39}$ erg s$^{-1}$ $\Msun$ yr$^{-1}$
from eq. (39) in \citet{Mineo_HMXB}.  Similarly our soft SED yields a 0.5--2 keV luminosity of $L_{\rm X, 0.5-2keV}^{\alpha=3} \approx 8 \times 10^{38}$ erg s$^{-1}$ $\Msun$ yr$^{-1}$,
 similar to the observed value of $L_{\rm X, 0.5-2keV}^{\rm obs, ISM} \approx 5 \times 10^{38}$ erg s$^{-1}$ $\Msun$ yr$^{-1}$
 from eq. (2) in \citet{Mineo_ISM}.  

We caution however that these comparisons are only illustrative.  
Moreover, it is difficult to guess how the relevant $z\sim15$ galaxies are different than local ones.  Using stacked spectra, \cite{Basu_Zych_2013} recently measured an evolution in the X-ray luminosity to SFR ratio for star-forming galaxies out to $z\sim4$.  On the theoretical side, there is also no reason to think that the first galaxies should have similar X-ray luminosities per SFR as local ones.  For example, the expected lower metallicity of early galaxies, discussed above in the context of intrinsic obscuration, could also result in a higher fraction of HMXBs (e.g. \citealt{Mirabel11, Fragos13}).
 Similarly, at these high redshifts Compton cooling might become important for the hot plasma (provided a large fraction of it manages to be blown out to low densities), thereby decreasing the fraction of its energy radiated away as soft X-rays. For Compton cooling to be more efficient than radiative cooling, the hot ISM (with $T\sim10^6$ K) needs to be at densities lower than: $\lsim 10^{-3}$-- $10^{-2}$ cm$^{-3}$ $[(1+z)/15]^4$, with the quoted range spanning solar to zero metallicities (e.g. \citealt{GS83}).  Modeling the ISM at these redshifts is very difficult, and model-dependent.  We note that numerical simulations of $\Tvir\approx10^4$ K galaxies at $z\sim$9--15 do find significant reservoirs of hot ISM at higher number densities (e.g. \citealt{Wise_Abel_2012}; Aykutalp et al., in prep), suggesting that radiative cooling would still dominate.  We postpone more detailed estimates to future work.
 Below we explore a wide range of $f_X$ values.

%\subsection{Summary of model parameters}
\vspace{+0.5cm}

To summarize, we have two fiducial models for the X-ray spectra of the first galaxies (c.f. the red and blue curves in Fig. \ref{fig:spectrum}): (i) a {\it hard} SED with a power-law index of $\alpha=0.8$, representing X-ray emission dominated by HMXBs; and (ii) a {\it soft} SED with a power-law index of $\alpha=3$, representing X-ray emission dominated by the hot ISM. 
In both cases, {\it metal-poor} gas inside the host galaxies is assumed to absorb all photons with energies below $E_0 = h \nu_0=$0.3 keV, corresponding to equivalent column densities as seen locally.
Finally, both SEDs are normalized according to their total number of X-ray photons escaping the galaxy per stellar baryon, with the fiducial choice being $f_X \equiv N_X/0.1 = 1$.

It is interesting to note that the combined observed SED from both components (hot ISM + HMXBs) in local galaxies strongly resembles our fiducial, $f_X=1$, hard ($\alpha=0.8$) SED for the first galaxies (see the red+blue curves in Fig. \ref{fig:spectrum}).  This is because the absorption of the hard HMXB power-law by a metal-enriched ISM is almost exactly compensated for by the absorbed soft contribution from the hot ISM. The resulting total emergent spectrum (in which the hot ISM and HMXBs comparably contribute to the soft-band luminosity per SFR) resembles an {\it unabsorbed}, $\alpha\sim1$ power law down to low energies.

%%%%%%%%%%%%%%%%%%%%%%%%%%%%%%%%%%%%%%%%%%%%%%%%%%%%%%%%%%%%%%%%%%%%%%
%% SECTION 3
%%%%%%%%%%%%%%%%%%%%%%%%%%%%%%%%%%%%%%%%%%%%%%%%%%%%%%%%%%%%%%%%%%%%%%
\section{Simulating the 21cm signal}
\label{sec:21cm}

The 21cm signal is usually represented in terms of the offset of the 21cm brightness temperature from the CMB temperature, $T_{\gamma}$, at an observed frequency $\nu$ \citep{FOB06}:
\begin{align}
\label{eq:delT}
\nonumber \delT(\nu) = &\frac{\Ts - \Tcmb}{1+z} (1 - e^{-\tau_{\nu_{21}}}) \approx \\
\nonumber &27 \nf (1+\delNL) \left(\frac{H}{dv_r/dr + H}\right) \left(1 - \frac{\Tcmb}{\Ts} \right) \\
&\times \left( \frac{1+z}{10} \frac{0.15}{\Omega_{\rm M} h^2}\right)^{1/2} \left( \frac{\Omega_b h^2}{0.023} \right) {\rm mK},
\end{align}
\noindent where $T_S$ is the gas spin temperature, $\tau_{\nu_{21}}$ is the optical depth at the 21cm frequency $\nu_{21}$, $\delNL({\bf x}, z) \equiv (\rho/\bar{\rho} - 1)$ is the evolved (Eulerian) density contrast, $H(z)$ is the Hubble parameter, $dv_r/dr$ is the comoving gradient of the line of sight component of the comoving velocity, and all quantities are evaluated at redshift $z=\nu_{21}/\nu - 1$.

To simulate the 21cm signal, we use a parallelized version of the publicly available \cmfast\footnote{http://homepage.sns.it/mesinger/Sim.html} code. 
It uses perturbation theory and excursion-set formalism to generate density, velocity, source, ionization, and spin temperature fields. For further details and tests of the code, interested readers are encouraged to see \cite{MF07}, \cite{MFC11}, and \cite{Zahn11}.
Here we outline our simulation set-up.

Our simulation boxes are 750 Mpc on a side, with a resolution of $500^3$.  Density fields are generated by perturbing Gaussian initial conditions, sampled on a higher-resolution, 2000$^3$ grid \citep{ZelDovich70}.  Ionization by UV photons is computed in an excursion-set fashion, by comparing the time-integrated number of ionizing photons to the number of neutral atoms in regions of decreasing scale \citep{FZH04}.  Specifically, a simulation cell at coordinate ${\bf x}$ is flagged as ionized if 
\begin{equation}
\label{eq:HII_barrier}
\zeta_{\rm UV} f_{\rm coll}({\bf x}, z, R, \Tvir) \geq 1-x_e({\bf x}, z, R) ~ ,
\end{equation}
\noindent where $\zeta_{\rm UV}$ is an ionizing efficiency parameter (here taken to be $\zeta_{\rm UV}=30$ so that our reionization histories match the mean observed value of the Thompson scattering optical depth to the CMB; \citealt{Hinshaw13}), and $f_{\rm coll}$ is the fraction of mass residing in dark matter halos with virial temperatures greater than $\Tvir$ inside a sphere of radius $R$ and mass $M=4/3 \pi R^3 \rho$, where $\rho = \bar{\rho} [1+\bar{\delta}_{\rm nl}]$, while $x_e$ is the fraction of gas partially-ionized by X-rays \citep{MFS13}.

We compute the Wouthuysen-Field (WF; \citealt{Wouthuysen52, Field58}) coupling (i.e. Ly$\alpha$ pumping; when the Ly$\alpha$ background from the first stars couples the spin temperature to the gas temperature) using the Lyman resonance backgrounds from both X-ray excitation of HI, and direct stellar emission.  The latter is found to dominate by two orders of magnitude for $f_X \sim 1$.  For the direct stellar emission, we assume standard Population II spectra from \citet{BL05_WF}, and sum over the Lyman resonance backgrounds \citep{MFC11}.

As reionization and WF coupling are not the focus of this work, we keep their relevant parameters fixed.  Instead we focus on the X-ray background, responsible for heating and partially-ionizing the IGM.  The angle-averaged X-ray specific intensity, $J(\nu, {\bf x}, z)$, (in erg s$^{-1}$ Hz$^{-1}$ cm$^{-2}$ sr$^{-1}$) can be computed integrating along the light-cone:
\begin{equation}
\label{eq:J}
J(\nu, {\bf x}, z) = \frac{(1+z)^3}{4\pi} \int_{z}^{\infty} dz' \frac{c dt}{dz'} \epsilon_{h \nu} ~ ,
\end{equation}
\noindent with the comoving specific emissivity evaluated at $\nu_e = \nu (1+z')/(1+z)$:
\begin{align}
\label{eq:emissivity}
\nonumber \epsilon_{h \nu}(\nu_e, {\bf x}, z') = &\alpha h \frac{N_{\rm X}}{\mu m_p} \left( \frac{\nu_e}{\nu_0} \right)^{-\alpha} \\
&\left[ \rho_{\rm crit, 0} \Omega_b f_\ast (1+\bar{\delta}_{\rm nl}) \frac{d f_{\rm coll}}{dt} \right] ~ ,
\end{align}
\noindent where $N_{\rm X}$ is the number of X-ray photons per stellar baryon (recall $f_X \equiv N_{\rm X}/0.1$), $\mu m_p$ is the mean baryon mass, $\rho_{\rm crit, 0}$ is the current critical density, $f_\ast$ is fraction of baryons converted into stars (we take $f_\ast=0.1$).  The IGM ionization and heating (including adiabatic and Compton) is tracked locally, with the X-ray contribution computed from eq. \ref{eq:J}, given the appropriate energy deposition fractions (taken from \citealt{FS10}).

Our fiducial model assumes that the ``first'' galaxies formed in atomically-cooled halos (with virial temperatures $\Tvir\geq10^4$ K)\footnote{The very first galaxies are likely hosted by smaller halos, in which gas accretes via the molecular cooling channel (e.g., \citealt{HTL96, ABN02, BCL02}).  However, H$_2$ is easily disrupted by external background radiation fields which sterilize star formation inside these ``mini-halos''.  Even just the background radiation from atomically-cooled halos is enough to sterilize mini-halos already at $z\gsim20$ (\citealt{HF12, MO12, Fialkov13}; Dijkstra et al. in prep).}, though we also consider a couple of models with $\Tvir\geq 10^5$ K, corresponding to inefficient star-formation in dwarf galaxies near the atomic cooling threshold.

%We focus on detecting the power spectrum in a single k-bin, centered around $k = 0.2 \, Mpc^{-1}$ (see \citealt{Pober13}, which showed that this scale is relatively free of foreground contaminations).

%To illustrate the relative impacts on the 21cm signal, we show the results of different runs, varying the virial temperature of the halo hosting galaxies, $T_{vir}$, theirX-ray production efficiencies $f_x$, and the SED of the source. The first two parameters control the relative time offset of reionization and IGM heating, while the last parameter is the one we focus our attention on. We describe them in turn.

%%%%%%%%%%%%%%%%%%%%%%%%%%%%%%%%%%%%%%%%%%%%%%%%%%%%%%%%%%%%%%%%%%%%%%
%% SECTION 4
%%%%%%%%%%%%%%%%%%%%%%%%%%%%%%%%%%%%%%%%%%%%%%%%%%%%%%%%%%%%%%%%%%%%%%
\section{Results}
\label{sec:results}

We take as the main observable the 3D 21cm power spectrum, defined as $P_{21}(k, z) = k^3/(2\pi^2 V) ~ \bar{\delT}(z)^2 \langle|\delta_{\rm 21}({\bf k}, z)|^2\rangle_k$, where $\delta_{21}({\bf x}, z) \equiv \delT({\bf x}, z)/ \bar{\delT}(z) - 1$.  We focus on the large-scale power, specifically at $k=0.2$ Mpc$^{-1}$.  These scales are small enough to be relatively clean of foregrounds \citep{Pober13}, while still large enough to achieve reasonable signal-to-noise (S/N) with even first generation instruments \citep{ME-WH14}. Our default power spectrum bin width is $d\ln k = 0.5$.  

\begin{figure}
\vspace{-1\baselineskip}
\hspace{-0.5cm}
\includegraphics[width=0.55\textwidth]{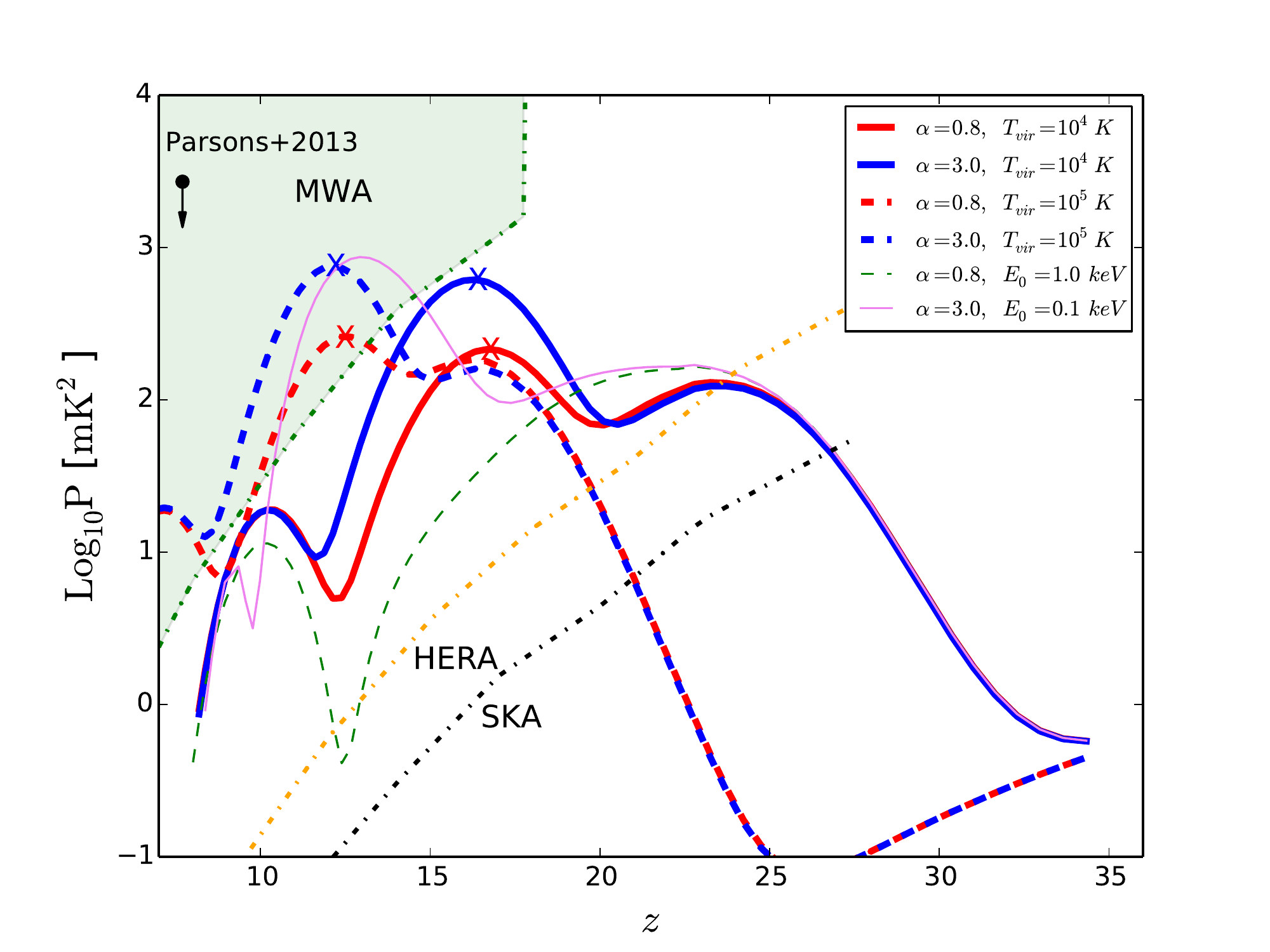}
\caption{Redshift evolution of the 21cm power amplitude at $k=0.2$ Mpc$^{-1}$.
%AM:new
The red and blue lines correspond to our two fiducial SEDs (soft and hard, respectively). The green (purple) line corresponds to an illustrative model with more (less) absorption by the host galaxy, such that all photons below $E_X \leq$ 1  (0.1) keV are absorbed locally inside the host's ISM.
%in which gas inside the first galaxies completely absorbs $E_X \leq 1 $ keV photons (also assuming $\Tvir=10^4$ K, $\alpha=0.8$, $f_X=1$). 
The three peaks in the fiducial models correspond to (from high to low redshift): (i) WF coupling, (ii) X-ray heating (labeled with an "X"), (iii) reionization.  We also show the 2$\sigma$ upper limit at $z\approx8$ from the PAPER experiment \citep{Parsons13}, as well as  $1\sigma$ thermal noise estimates corresponding to a 1000h observation with some upcoming and current instruments (taken from \citealt{ME-WH14}).}
\label{fig:z_evolution}
\end{figure}

In most models, the amplitude of the large-scale 21cm power rises and falls with three prominent peaks (e.g. \citealt{PF07, Baek10}; see also Fig. \ref{fig:z_evolution}), corresponding to (from high to low redshift): (i) WF coupling; (ii) X-ray heating; (iii) reionization.  These are sourced mainly by fluctuations in the (i) WF coupling coefficient (sourced by the \lya\ background); (ii) gas temperature; (iii) ionized fraction, respectively.  WF coupling and reionization are most likely dominated by direct stellar emission of UV photons \citep{MO12, MFS13}.  Hence, here we focus on the middle, ``X-ray heating'' peak in 21cm power, shortly after the X-rays from the first galaxies started heating the cosmic gas, and the spatial fluctuations in gas temperature\footnote{In most models we consider, the spin temperature, $T_S$, is already closely coupled to the gas kinetic temperature at the time of the peak in 21cm power associated with heating.  Hence we use the two terms interchangeably below.} were the largest.  After $T_S \gg T_\gamma$, the 21cm signal is no longer sensitive to the thermal state of the gas (c.f. eq. \ref{eq:delT}).

%% FIGURE 1 %%%%%%%%%%%%%%%%%%%%%%%%%%%%%%%%%%%%%%%%%%%%%%%%%%%%%%%%%%%%%%%%%
\begin{figure}
\begin{minipage}[c][19cm][t]{0.5\textwidth}
  \centering
  \includegraphics[width=1\textwidth]{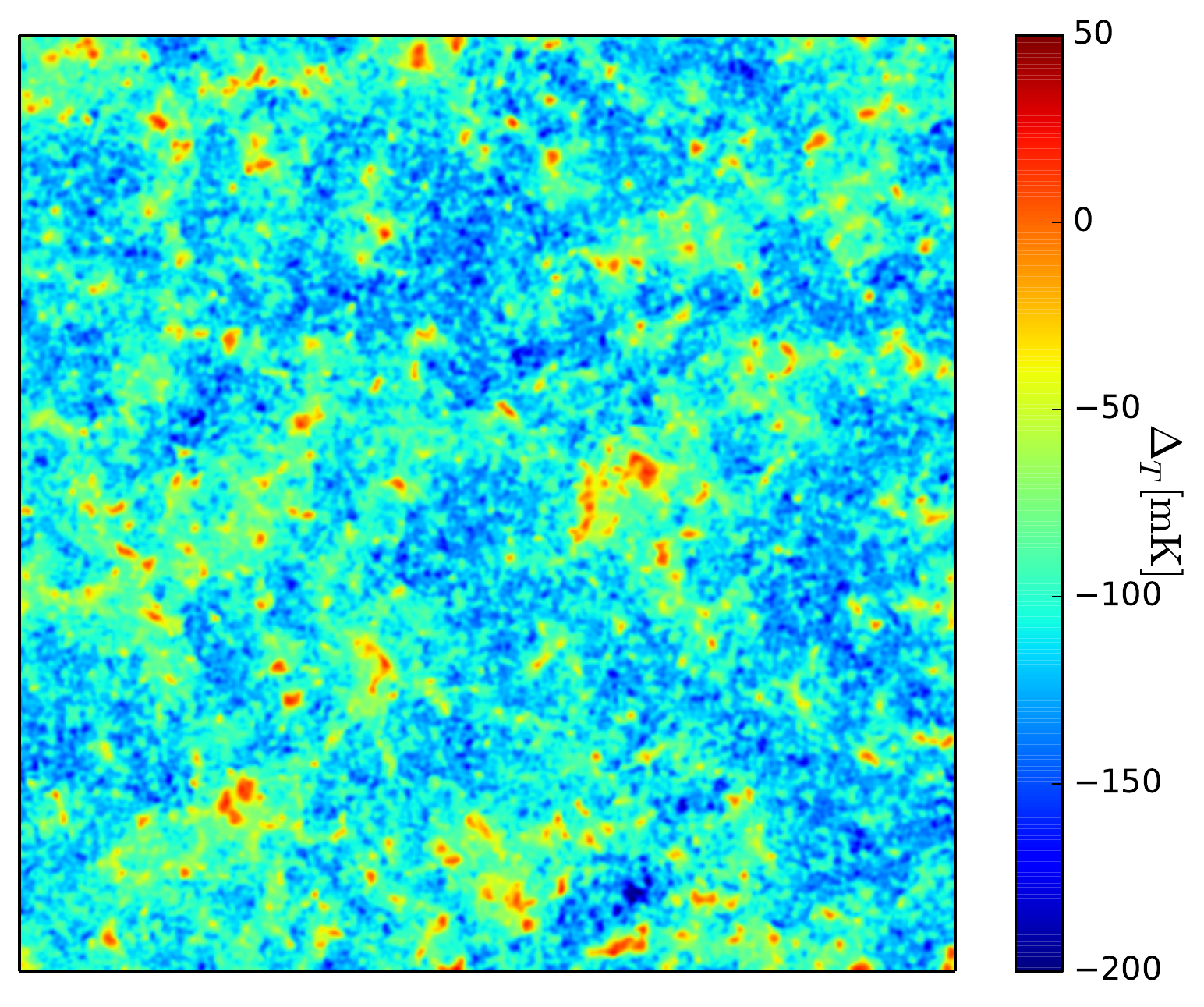} 
  \label{fig:slice_0.8}\par\vfill
  \includegraphics[width=1\textwidth]{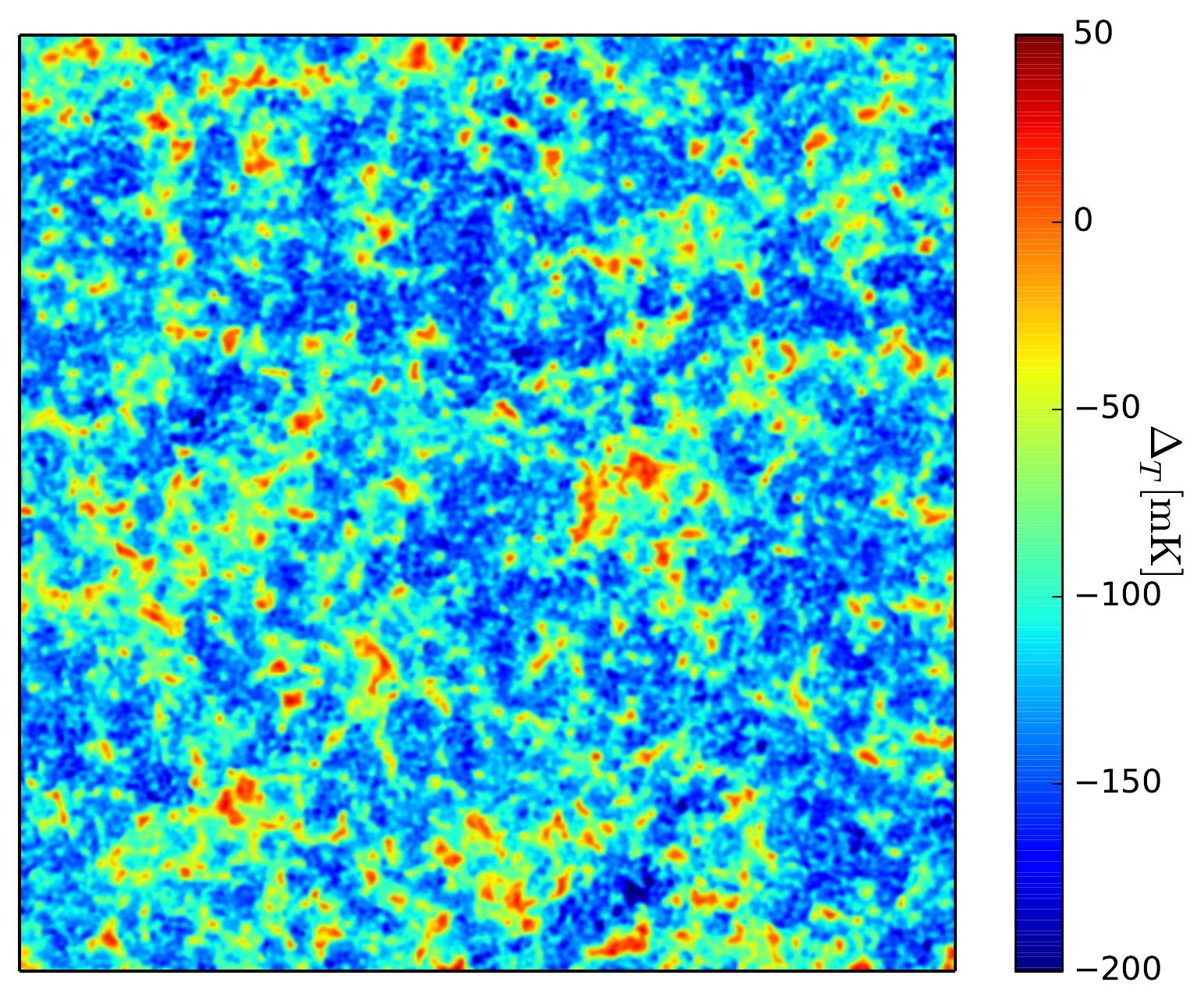}
  \label{fig:slice_3.0}
\caption{Slices through the 21cm brightness temperature fields for our fiducial model with a hard (top) and soft (bottom) X-ray SED.  Slices are 1.5 Mpc thick, and are taken at $\zpeak=16.7$ (top) and $\zpeak=16.3$ (bottom), corresponding to the redshift when the $k = 0.2$ Mpc$^{-1}$ 21cm power is the largest for each model.
}
\label{fig:slices}
\end{minipage}
\end{figure}
%%%%%%%%%%%%%%%%%%%%%%%%%%%%%%%%%%%%%%%%%%%%%%%%%%%%%%%%%%%%%%%%%%%

In Fig. \ref{fig:slices} we show slices through the 21cm brightness temperature maps for our fiducial model ($f_x=1$, $\Tvir=10^4$ K), with the hard (soft) SED at the top (bottom).  Both of the slices are taken at roughly the same redshift [$\zpeak=$16.7 (16.3) for the top (bottom) panels], when the large-scale ($k = 0.2$ Mpc$^{-1}$) 21cm power is the strongest in each model, corresponding to the X-ray heating peak.  It is evident from the figure that the hard SED results in more uniform brightness temperature maps, due to the longer distances traveled by more energetic photons.

%% FIGURE 2 %%%%%%%%%%%%%%%%%%%%%%%%%%%%%%%%%%%%%%%%%%%%%%%%%%%%%%%%%%%%%%%%%
\begin{figure}
\vspace{-1\baselineskip}
\hspace{-0.5cm}
\includegraphics[width=0.55\textwidth]{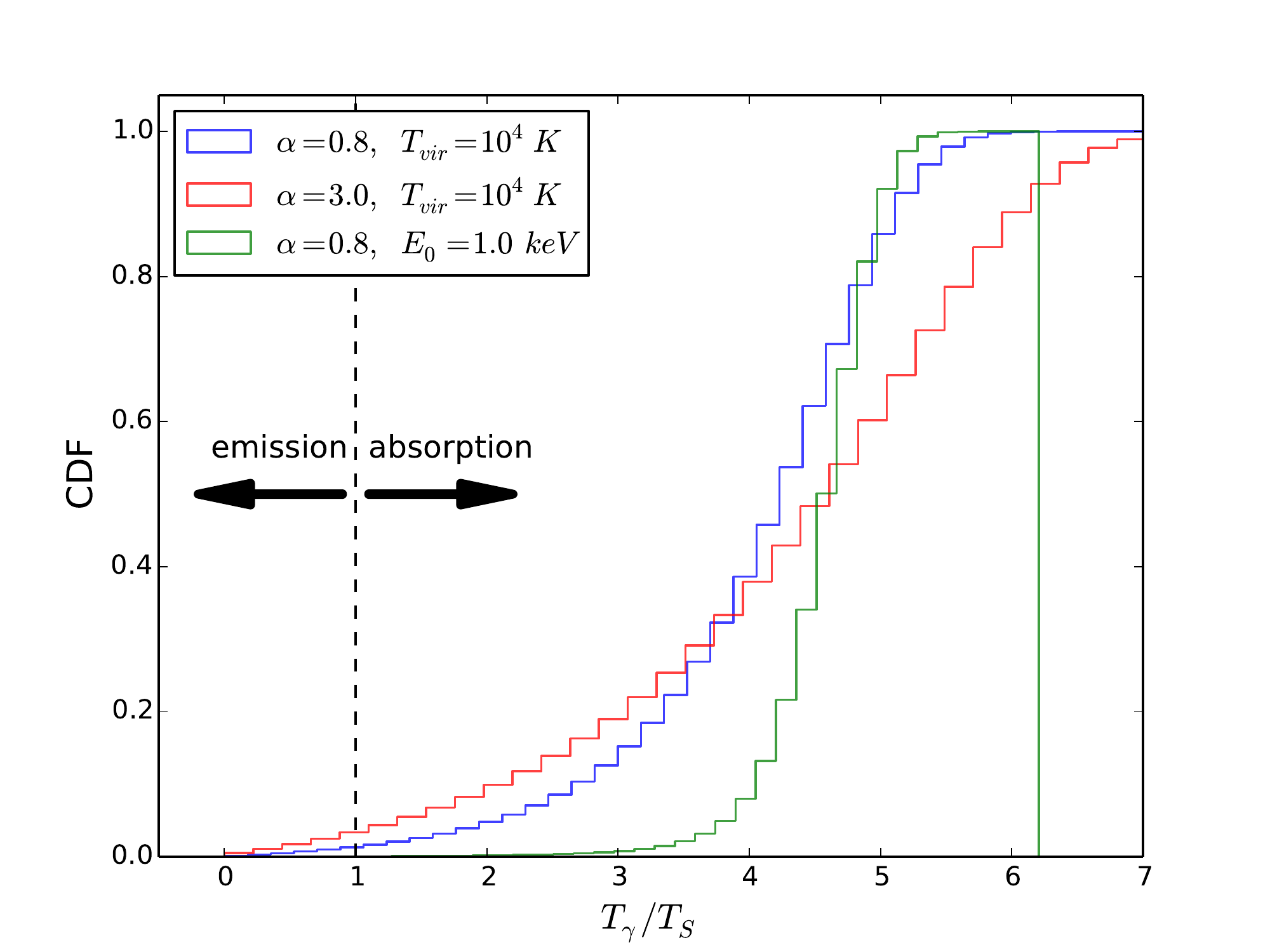}
\caption{CDFs of $T_{\gamma}/T_S$ corresponding to the fiducial soft/hard SED models shown in Fig. \ref{fig:slices}. With the green curve, we also show an extremely hard SED model, in which all soft photons with energies below $E_0 = 1$ keV are assumed to be absorbed by their host galaxies.  The curves correspond to $z\sim16.5$.}
\label{fig:T_distributions}
\end{figure}
%%%%%%%%%%%%%%%%%%%%%%%%%%%%%%%%%%%%%%%%%%%%%%%%%%%%%%%%%%%%%%%%%%%%%%

We quantify this in Fig. \ref{fig:T_distributions} which shows the corresponding temperature distributions.  Specifically, we plot the cumulative distribution functions (CDFs) of $T_{\gamma}/T_S$ (c.f. eq. \ref{eq:delT}) for the soft (red) and hard (blue) SEDs.  The soft SED has a noticeably broader distribution of temperatures.  Because heating is more patchy in this model, there are large regions distant from galaxies which are still cooling adiabatically at this epoch.  Such cold gas is absent in the model with the hard SED.  With the green curve we also show an extremely hard SED model, in which all soft photons with energies below 1 keV are assumed to be absorbed by their host galaxy.  With no soft photons, the heating is extremely uniform, with the temperature distribution approaching a step function.

%% FIGURE 3 %%%%%%%%%%%%%%%%%%%%%%%%%%%%%%%%%%%%%%%%%%%%%%%%%%%%%%%%%%%%%%%%%
\begin{figure}
\vspace{-1\baselineskip}
\hspace{-0.5cm}
\includegraphics[width=0.55\textwidth]{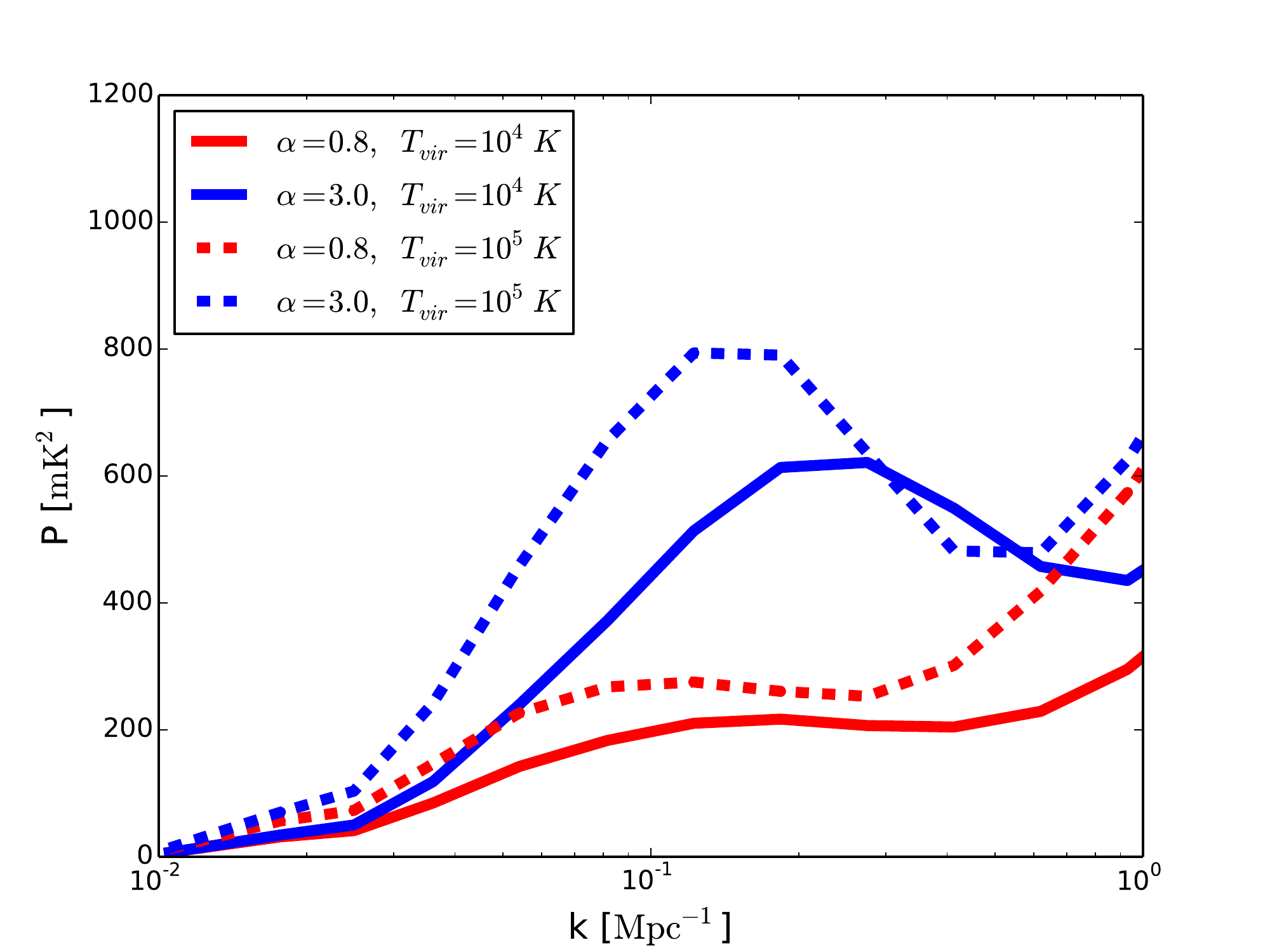}
\caption{Power spectra corresponding to soft (blue curves) and hard (red curves) SEDs.  Galaxies are assumed to be hosted by halos with virial temperatures greater than 10$^4$ K (solid curves) and 10$^5$ K (dotted curves). As above, power spectra are taken at the redshift where the amplitude at $k=0.2$ Mpc$^{-1}$ is the largest, corresponding to $\zpeak\sim16.5$ and 12, for the $\Tvir=10^4$ K and 10$^5$ K models, respectively.}
\label{fig:power_peak}
\end{figure}
%%%%%%%%%%%%%%%%%%%%%%%%%%%%%%%%%%%%%%%%%%%%%%%%%%%%%%%%%%%%%%%%%%%%%%

During the X-ray heating epoch, the temperature distribution sets the 21cm power.  Broader distributions generally result in higher power spectrum amplitudes.  The 21cm power spectra corresponding to the fiducial soft/hard SED models from Fig. \ref{fig:slices} are shown in Fig. \ref{fig:power_peak}.  Indeed the 21cm power is higher with the soft SED, corresponding to X-rays from the hot ISM.  On large scales, this difference is a factor of $\sim$ 3. Interestingly, the power-spectrum peaks at a scale of $k\sim0.2 \, \mathrm{Mpc}^{-1}$, which corresponds to the mean free path of the average photon in this model, $E_X\sim 0.5$ keV.  We caution however that more realistic models for the intrinsic absorption of the host galaxies could smear out this feature.

In Fig. \ref{fig:power_peak} we also include power spectra for the same fiducial soft/hard SEDs, but assuming instead that the dominant galaxy population corresponds to much more massive systems, with $\Tvir\geq 10^5$ K.  If such massive systems are required for efficient star-formation, the X-ray heating peak is delayed until $\zpeak\sim12$ (see Fig. \ref{fig:z_evolution}).
These host halos are more biased than those corresponding to our fiducial, $\Tvir=10^4$ K ones.  The additional modulation by their larger correlation lengths drives the power spectrum peak to somewhat larger scales, $k\sim0.1 \, \mathrm{Mpc}^{-1}$.  Interestingly, the large-scale power in the soft SED model is also a factor of $\sim$ 3 higher than in the hard SED model, as was the case for our fiducial choice of $\Tvir$; we elaborate more on this below.

%% FIGURE 4 %%%%%%%%%%%%%%%%%%%%%%%%%%%%%%%%%%%%%%%%%%%%%%%%%%%%%%%%%%%%%%%%%
%%%%%%%%%%%%%%%%%%%%%%%%%%%%%%%%%%%%%%%%%%%%%%%%%%%%%%%%%%%%%%%%%%%%%%

From here on, we focus only on the amplitude of the 21cm power at $k=0.2$ Mpc$^{-1}$, lying in the narrow $k$-space window accessible with the first generation interferometers, as mentioned above.  We note however that the evolution of large-scale power is qualitatively self-similar for a wide range of wave-numbers (e.g. \citealt{Baek10, Santos11}).  In Fig. \ref{fig:z_evolution} we show the redshift evolution of the 21cm power amplitude at $k=0.2$ Mpc$^{-1}$.  The aforementioned characteristic three-peaked structure is evident, with the WF coupling, X-ray heating and reionization peaks clearly separated in the fiducial models, assuming either $\Tvir=10^4$ or 10$^5$ K.  In the figure we also present S/N estimates for three interferometric arrays: the MWA \citep{Beardsley12}, the proposed second-generation Hydrogen Epoch of Reionization Arrays (HERA; http://reionization.org/; Pober et al., in prep), and the SKA Phase 1 \citep{Dewdney13}.  The S/N is computed assuming a 1000h integration, with a fixed $\Delta z = 0.5$ band.  The MWA is currently taking data, and is expected to achieve 1000h integration in $\gsim2016$, while HERA and the SKA are expected to begin taking observations in a few years.  We note that a 1000h survey with SKA Phase 1 is planned as part of the early science results (Koopmans et al. in prep).  For details on the noise calculations please see \citet{ME-WH14}.

Here we again see explicitly that the X-ray SED impacts the X-ray heating epoch (i.e. the middle peak). 
% If the soft X-ray emission from the first galaxies was dominated by the hot ISM, the X-ray heating epoch would be detectable at S/N$\gsim$ few even with the first-generation MWA.
  Moreover, the difference between the soft and hard SEDs is comparable (factor of $\sim3$ in power amplitude at the X-ray heating peak) for two very different choices of $\Tvir$.

In Fig. \ref{fig:z_evolution} we also include the redshift evolution in the $E_0=1$ keV model (also assuming $\Tvir=10^4$ K, $\alpha=0.8$, $f_X=1$), corresponding to host-galaxy intrinsic absorption which is considerably stronger than what is observed in local galaxies.  The high energy ($\gsim$ 1 keV) X-rays which manage to escape the first galaxies in this model, heat the IGM uniformly and inefficiently (as the absorption cross-section is smaller).  The uniformity of heating dramatically suppresses the temperature fluctuations (see Fig. \ref{fig:T_distributions}), and therefore there is no noticeable X-ray heating peak (see also \citealt{MFS13}).  Furthermore, these high energy photons interact weakly with the IGM, only managing to heat it to temperatures $T_S \sim T_\gamma$ when reionization commences.  The resulting narrow temperature distribution centered around $(1-T_\gamma/T_S) \sim 0$ (c.f. eq. \ref{eq:delT}) results in a much deeper drop in power at $z\sim12$.  Even at the midpoint of reionization ($z\sim10$), the neutral IGM patches are only heated to $(1-T_\gamma/T_S) \sim$ 0.6--0.7, which suppresses the amplitude of the 21cm signal by tens of percent.  

We stress that the $E_0=1$ keV model is quite extreme because we observe copious low-energy X-rays escaping even from nearby galaxies, which are more massive and evolved than the first galaxies.  Instead, it is more likely that the first galaxies had even lower host galaxy absorption, owing to lower average HI column densities (as discussed above).  To illustrate such a scenario, we also include i Fig. \ref{fig:z_evolution} a model with $E_0=0.1$ keV, shown with a thin purple curve (also assuming $\Tvir=10^4$ K, $\alpha=3$, $f_X=1$).  The X-ray peak is not significantly increased in such a model, as even our fiducial hot ISM SED is soft enough to leave standing large cold patches when the IGM near the galaxies is heated (see Fig. \ref{fig:T_distributions}).  However, the peak location is shifted to lower redshifts, since the mean energy absorbed by the IGM is considerably lower in this model.

%  Nevertheless it serves to highlight the importance of the intrinsic absorption in determining the 21cm signal.

\subsection{Is the imprint of the SED degenerate with the luminosity?}

%% FIGURE 5 %%%%%%%%%%%%%%%%%%%%%%%%%%%%%%%%%%%%%%%%%%%%%%%%%%%%%%%%%%%%%%%%%
\begin{figure}
\vspace{-1\baselineskip}
\hspace{-0.5cm}
\includegraphics[width=0.55\textwidth]{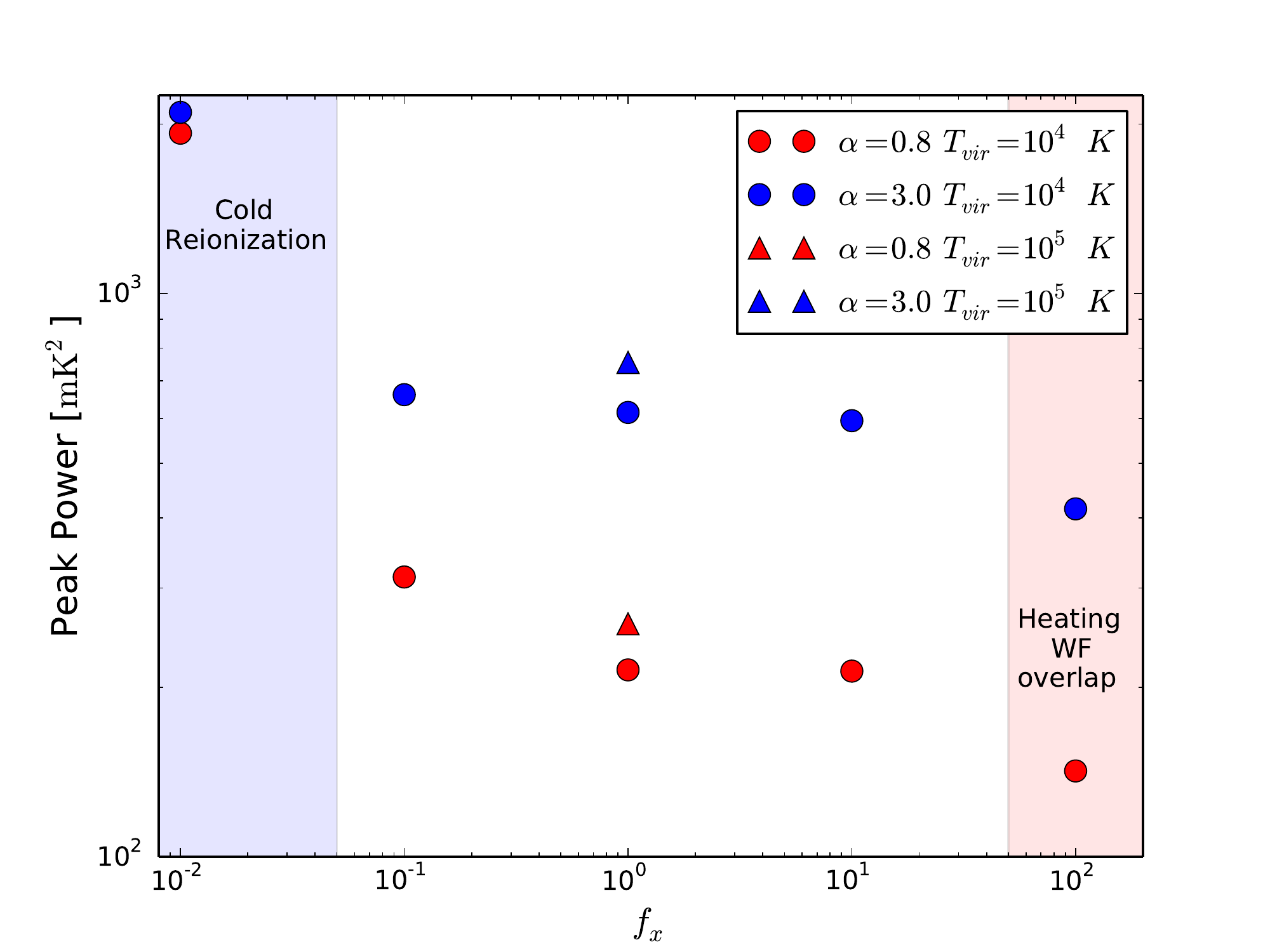}
\caption{Peak amplitude (over all redshifts) of the $k=0.2$ Mpc$^{-1}$ 21cm power, as a function of $f_X$.  Circles (triangles) correspond to models with $\Tvir= 10^4$ (10$^5$) K. Red (blue) colors indicate hard (soft) SEDs.  The regions of parameter space in which the X-ray heating epoch overlaps with reionization and WF coupling are shaded in blue and pink, respectively.}
\label{fig:power_vs_fx}
\end{figure}
%%%%%%%%%%%%%%%%%%%%%%%%%%%%%%%%%%%%%%%%%%%%%%%%%%%%%%%%%%%%%%%%%%%%%%

%% FIGURE 6 %%%%%%%%%%%%%%%%%%%%%%%%%%%%%%%%%%%%%%%%%%%%%%%%%%%%%%%%%%%%%%%%%
\begin{figure}
\vspace{-1\baselineskip}
\hspace{-0.5cm}
\includegraphics[width=0.55\textwidth]{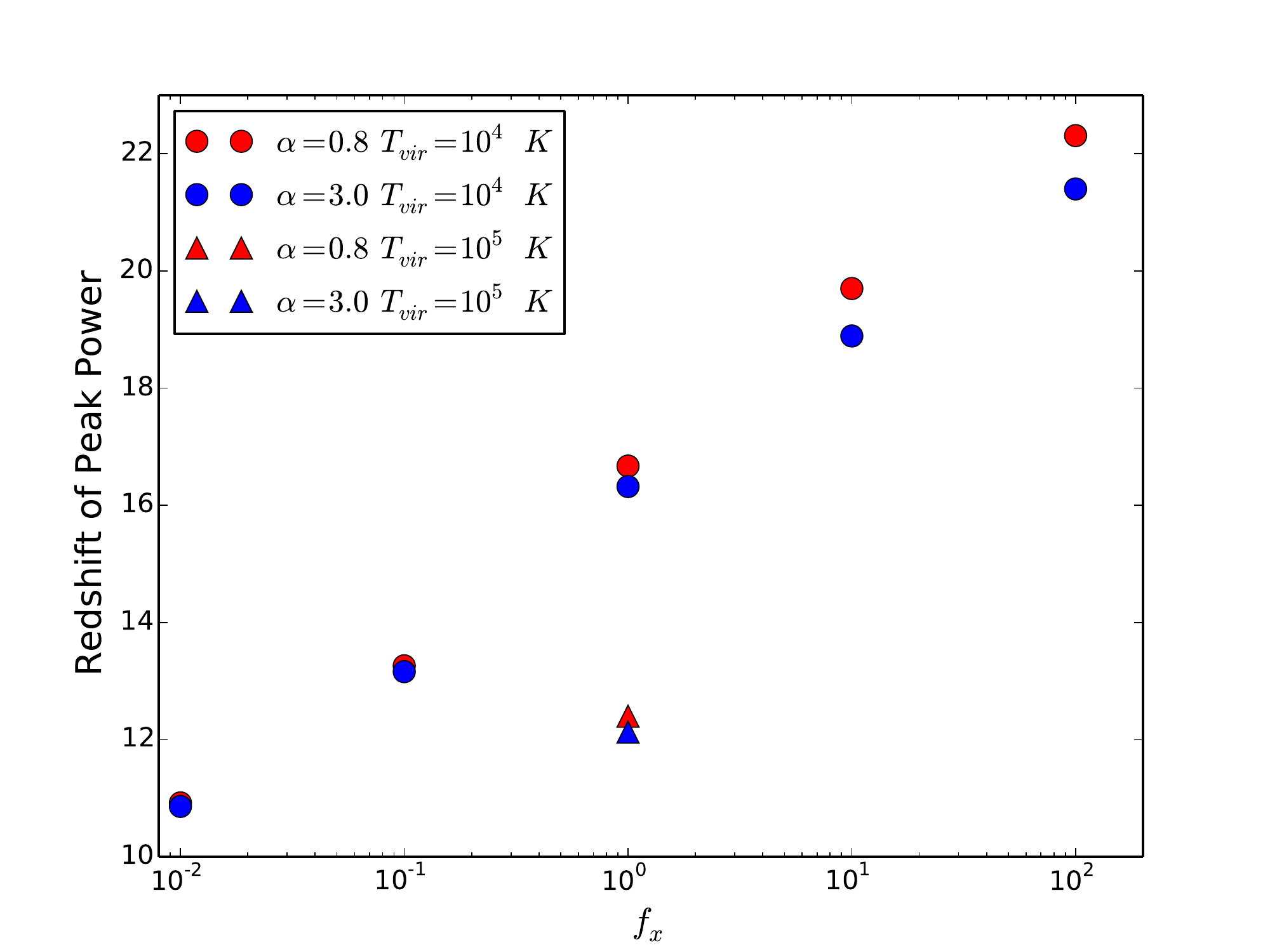}
\caption{Same as Fig. \ref{fig:power_vs_fx}, but instead showing the redshift at which the $k=0.2$ Mpc$^{-1}$ 21cm power peaks on the y-axis.}
\label{fig:zpeak_vs_fx}
\end{figure}
%%%%%%%%%%%%%%%%%%%%%%%%%%%%%%%%%%%%%%%%%%%%%%%%%%%%%%%%%%%%%%%%%%%%%%

We noted above that the difference in the amplitude of the X-ray heating peaks for soft vs hard SEDs was remarkably similar even assuming very disparate values of $\Tvir$.  In Fig. \ref{fig:power_vs_fx} we further investigate the robustness of the SED imprint by plotting the peak amplitude of the $k=0.2$ Mpc$^{-1}$ 21cm power, as a function of $f_X$.  We remind the reader that $f_X$ is a proxy for the total number of X-ray photons per stellar baryon, serving to normalize our SEDs.

We explore a wide range of values: $10^{-2} < f_X < 10^2$.  {\it For a given SED}, the amplitude of the peak power is remarkably constant over a wide range of luminosities: $10^{-1.5} \lsim f_X \lsim 10^{1.5}$, as already noted in \citet{ME-WH14}.  This is due to the fact that the large-scale 21cm power peaks when the large-scale temperature fluctuations are maximized, which is roughly a self-similar process occurring at $\bar{\delta T}_b \sim -100$ mK \citep{ME-WH14}.  If $f_X \lsim 10^{-1.5}$, the first galaxies are too faint in X-rays to heat the IGM before reionization commences.  The resulting contrast between the ionized and very cold neutral patches can drive up the 21cm signal considerably (e.g. \citealt{Parsons13}).  On the other hand if $f_X \gsim 10^{1.5}$, X-ray heating merges with WF coupling, and the spin temperature is never able to fully couple to the kinetic temperature, $T_K$, before the gas was heated.  If $T_S > T_K$ during the X-ray heating peak, the amplitude of the signal decreases.

On the other hand, the amplitude of the peak power {\it is} strongly affected by the assumed SED.  For our soft SED, corresponding to the hot ISM, the 21cm signal peaks at $P_{21}\sim$ 600-700 mK$^2$ over a broad range of $f_X$ and $\Tvir$ values.  Similarly, for our hard SED, corresponding to HMXBs, the 21cm signal peaks at $P_{21}\sim$ 200-300 mK$^2$, again over a broad range: $10^{-1.5}\lsim f_X\lsim 10^{1.5}$ and $10^4$ K $\lsim \Tvir\lsim$ 10$^5$ K.  Therefore, {\it the peak amplitude of the large-scale 21cm power is a robust probe of the X-ray SED of the first galaxies}, not degenerate with their X-ray luminosities and host halo masses.

Finally, Fig. \ref{fig:zpeak_vs_fx} shows the redshift of the peak power, $\zpeak$, on the vertical axis, as a function of the X-ray efficiency $f_x$. An increase in the X-ray efficiency, or a decrease in the host halo virial temperature, corresponds to a higher $\zpeak$, since the IGM is heated earlier.
  However, the X-ray SED does not have a strong impact on the redshift of the peak power.

Thus, upcoming interferometers can determine the X-ray luminosities and host halo mass of the first galaxies from the redshift of the peak 21cm power \citep{ME-WH14}, while the {\it amplitude of the peak power can constrain the SED}.  Hence we should soon have a complete picture of the X-ray properties of the first galaxies.

%%%%%%%%%%%%%%%%%%%%%%%%%%%%%%%%%%%%%%%%%%%%%%%%%%%%%%%%%%%%%%%%%%%%%%
%% SECTION 5
%%%%%%%%%%%%%%%%%%%%%%%%%%%%%%%%%%%%%%%%%%%%%%%%%%%%%%%%%%%%%%%%%%%%%%

\section{Conclusions}
\label{sec:conc}

In this proof-of-concept, we investigated if the X-ray SED of the first galaxies could have a robust imprint in the 21cm signal.  We were motivated by {\it Chandra} observations of local star-forming galaxies, in which the relevant soft-band luminosity has two main contributors: (i) the hot, diffuse ISM and (ii) HMXBs.  Using simple SEDs corresponding to these two populations, we studied their imprint on the 21cm signal, focusing on the epoch when the first galaxies began heating the IGM with their X-rays.

Understandably a soft SED, representative of the hot ISM, results in larger fluctuations of the IGM temperature, as the absorption cross-section for X-rays has a strong dependence on the photon energy.  Low energy photons are much more likely to be absorbed closer to the galaxies.  These stronger temperature fluctuations drive up the amplitude of the large-scale ($k\sim0.2$ Mpc$^{-1}$) 21cm power by a factor of $\sim$ 3, compared with models dominated by hard SEDs representative of HMXBs.

More generally, we show that the X-ray SED determines the amplitude of the peak 21cm power,
for a wide range of X-ray luminosities ($10^{-1.5}\lsim f_X\lsim 10^{1.5}$) and host halo virial temperatures ($10^4$ K $\lsim \Tvir\lsim$ 10$^5$ K). The reverse is true for the redshift at which the large-scale power peaks: it is insensitive to the SED, and instead determined by the X-ray luminosity and host halo mass of the first galaxies.   Thus, upcoming interferometers can determine the X-ray luminosities and host halo mass of the first galaxies from the redshift of the peak 21cm power, while the amplitude of the peak power will constrain the SED.

The absorption intrinsic to the host galaxies remains a significant source of uncertainty, and could substantially impact the emerging soft-band SED.  In this work, our fiducial models assume a sharp cut below $E_0 = 0.3$ keV, corresponding to photons whose mean free path is greater than unity given the same column densities observed in local galaxies, {\it but} assuming a low metallicity such that the ISM absorption is dominated by hydrogen and helium.  If on the other hand the emerging X-ray SED of the first galaxies is {\it exactly} as observed in local ones, the 21cm signal should be as predicted in our hard, $\alpha=0.8$ models.  This is because in observed composite spectra (see Fig. \ref{fig:spectrum}), the additional absorption of the HMXB template at $\lsim1$ keV provided by the metals is almost exactly compensated for by the additional contribution from the absorbed, hot ISM.  This makes the total spectrum (absorbed hot ISM + absorbed HMXBs) look like an unabsorbed, $\alpha\sim0.8$ power law down to low energies ($\lsim 0.5$ keV), which is effectively our hard SED for the first galaxies.

Subsequent work will focus on constructing more complete and detailed models of the X-ray SEDs emerging from the first galaxies, guided by hydrodynamic simulations including metal pollution.  Combined with upcoming 21cm interferometric observations, we will be able to robustly study high energy processes inside the first galaxies.

\vspace{+1cm}
We thank Bret Lehmer for stimulating conversations which contributed to motivating this work.
%AM: send to bret, jonathan+?, leon for comments..
SM acknowledges support from the NASA's Astrophysics Data Analysis Program (ADAP) grant NNH13CH56C.

%%%%%%%%%%%%%%%%%%%%%%%%%%%%%%%%%%%%%%%%%%%%%%%%%%%%%%%%%%%%%%%%%%%%%%
%%  REFERENCES
%%%%%%%%%%%%%%%%%%%%%%%%%%%%%%%%%%%%%%%%%%%%%%%%%%%%%%%%%%%%%%%%%%%%%% 

\bibliographystyle{mn2e}
\bibliography{ms}

\end{document}